\documentclass{mn2e}
\usepackage{times}
\input{psfig.sty}
\newif\ifAMStwofonts

\title{Constraints on jet X-ray emission in low/hard state X-ray binaries}

\author[Maccarone] {Thomas J. Maccarone\\ Astronomical Institute
``Anton Pannekoek'', University of Amsterdam, Kruislaan 403, 1098 SJ,
Amsterdam, The Netherlands}

\date{}

\begin{document}

\maketitle

\label{firstpage}

\def\simlt{\mathrel{\rlap{\lower 3pt\hbox{$\sim$}}
        \raise 2.0pt\hbox{$<$}}}
\def\simgt{\mathrel{\rlap{\lower 3pt\hbox{$\sim$}}
        \raise 2.0pt\hbox{$>$}}}

\input epsf

\begin{abstract}
We show that the combination of the similarities between the X-ray
properties of low luminosity accreting black holes and accreting
neutron stars, combined with the differences in their radio properties
argues that the X-rays from these systems are unlikely to be formed in
the relativistic jets.  Specifically, the spectra of extreme island
state neutron stars and low/hard state black holes are known to be
indistinguishable, while the power spectra from these systems are
known to show only minor differences beyond what would be expected
from scaling the characteristic variability frequencies by the mass of
the compact object.  The spectral and temporal similarities thus imply
a common emission mechanism that has only minor deviations from having
all key parameters scaling linearly with the mass of the compact
object, while we show that this is inconsistent with the observations
that the radio powers of neutron stars are typically about 30 times
lower than those of black holes at the same X-ray luminosity.  We also
show that an abrupt luminosity change would be expected when a system
makes a spectral state transition from a radiatively inefficient jet
dominated accretion flow to a thin disk dominated flow, but that such
a change is not seen.

\end{abstract}


\section{Introduction}
Numerous similarities are have been found between low magnetic field
accreting neutron stars and accreting stellar mass black holes.  These
similarities are most striking when the systems are accreting at
relatively low fractions of the Eddington luminosity, $L_{EDD}$ --
below the few percent of the Eddington limit where spectral state
transitions from soft, quasi-thermal to hard power law dominated
spectra are typically seen (Maccarone 2003), but above the
$\sim10^{-5} L_{EDD}$ where the thermal emission from the surface of
the neutron star becomes a substantial fraction of its total
luminosity.  In this range, the X-ray spectral energy distributions of
black holes and neutron stars show no notable differences, and the
X-ray power spectra show only minor differences apart from those which
are attributable to the smaller mass and hence smaller characteristic
timescale for the neutron stars.  The similarity between the X-ray
properties of these low magnetic field neutron stars and those of
black holes strongly supports the suggestion that the fundamental
physical processes in these systems producing X-ray emission in these
systems must also be quite similar.  On the other hand, the radio
properties of accreting neutron stars and accreting black holes are
quite different; at the same X-ray luminosity, the neutron stars are
typically about 30 times fainter than the black holes (Fender \&
Hendry 2000; Migliari et al. 2003).  In this paper, we will show that
the similarity between the X-ray properties of neutron stars and black
holes combined with the differences between their radio properties
places strong constraints on the possibility that scale invariant jets
can be responsible for the X-ray emission in both neutron stars and
accreting black holes.  Instead, these results provide support for the
idea that the X-rays are, indeed, produced in a disk plus hot coronae
system, and that the efficiency in extracting the disk's energy into
the jet is different for black holes and neutron stars.  We also show
that the smooth changes in luminosity across transition between the
different spectral states are at odds with the predictions of jet
models that the radiative efficiency of the accretion flow should
change dramatically at the state transitions.

\section{Competing pictures for X-ray binary spectral modeling}
\subsection{Black hole X-ray spectra}
Black hole X-ray binaries are found in several spectral states, the
two most common being the low/hard state, usually found at
luminosities of less than about 2\% of the Eddington limit and the
high/soft state, usually found at luminosities between about 2 and
30\% of the Eddington limit. Above 30\% of the Eddington limit,
sources are usually in the very high state.  We refer the reader to
Nowak (1995) for a summary of the spectral state phenomenology.  We
also note that there is not one-to-one correspondence between
luminosity and spectral state (see e.g. Miyamoto et al. 1995; Homan et
al. 2001; Maccarone \& Coppi 2003a - MCa).

We will focus on the two lower luminosity states, the low/hard state
and the high/soft state.  The high/soft state energy spectrum is
typically very well fit by multi-temperature blackbody models, often
with a weak power law tail. The high/soft state thus represents the
case where the accretion flow is that of a standard Shakura-Sunyaev
(1973) geometrically thin, optically thick disk.  The low/hard state
spectra fit well phenomenologically to cutoff power laws, with a
photon spectral index of about 1.7 and a cutoff at about 100 keV.
These have historically been modeled as seed photons being Compton
upscattered in a thermal plasma with a temperature of about 100 keV,
and an optical depth of about unity (see e.g. Pottschmidt et
al. 2003).  More recently, it has been suggested that these low/hard
state spectra can also be modeled as sychrotron X-ray jets (Markoff,
Falcke \& Fender 2001).  In both spectral states, broad spectral bumps
are seen at energies of about 30 keV, in addition to some lines edges
from various elements.  These features are generally attributed to
reflection of the hard power law photons off the accretion disk
(e.g. George \& Fabian 1991).

It has been argued that some of the observed features are rather
difficult to reproduce in a jet model -- most notably the cutoff at
about 100 keV and the evidence of reflection off the disk, which
should not be seen if the X-ray emitting region is beamed strongly
away from the disk (Poutanen \& Zdziarski 2002).  The former point
boils down to the fact that the cutoff energies in X-ray binaries are
almost always found to be very close to 100 keV, while in blazars, the
synchrotron cutoff energy varies by large factors. However,
measurements of the spectral cutoff in the $\gamma$-rays are made over
a dramatically smaller range in luminosity than measurements of blazar
cutoff energies, and the velocities of the X-ray emission regions of
the jets are typically less than 0.3$c$.  Heinz (2005) showed that
pure synchrotron jets would have trouble simultaneously reproducing
the observed X-ray spectra of X-ray binaries and AGN along with
observed correlations between their masses and their fluxes in radio
and X-rays (Merloni, Heinz \& Di Matteo 2003; Falcke, K\"ording \&
Markoff 2004), but could not rule out the case that inverse Compton
emission from jets might be important.  More stringent tests of the
jet X-ray model are thus needed.

\subsection{Neutron star X-ray spectra}
The spectral state nomenclature for neutron stars is a bit different
than that for black holes.  Furthermore, the analogies between the
states of neutron stars and the states of black holes have recently
shown the need for refinement.  The relatively low luminosity neutron
stars (i.e. the atoll sources) were originally divided into island
states, similar to the low hard states of black holes, and banana
states, similar to the high/soft states (Hasinger \& van der Klis
1989).  More recently, the need for ``extreme island states'' (EIS)
has been shown, in the sense that the island state properties
correspond to something intermediate between the low/hard and high
soft states of black holes, while the EIS corresponds much more
closely to the low/hard state of black holes (see e.g. van der Klis
2005 and references within).  When these EIS neutron stars' spectra
are examined, they are found to be quite similar to those of black
holes in the low/hard state -- compare for example the spectral fit
parameters for Cygnus X-1 in Pottschmidt et al. (2003) with those of
the neutron star Aql X-1 in its extreme island state in Maccarone \&
Coppi (2003b - MCb) -- in both cases, the spectra are fit well by
thermal Comptonization models with optical depth of about 1 and
temperature of about 100 keV.  An incorrect lore has developed that
neutron stars never show spectra as hard as the low/hard state black
holes.  This is most likely because of the old belief that the
low/hard state corresponded not to the EIS, but the classical island
state.  In the classical island state, typically optical depths of a
few and electron temperatures of order 20 keV are found from
Comptonization model fits (MCb), yielding softer spectra than low/hard
state black holes.

\section{Similarities of variability properties in low/hard states}
The timing signatures of both low/hard state and EIS systems are also
very similar, being well described by a series of broad Lorentzians
(see Nowak 2000 for a typical black hole; van Straaten et al. 2002 for
a typical neutron star), with three key exceptions -- the first being
that the typical frequencies in the neutron star power spectra are a
factor of about 5 lower than in black holes, the second being that the
neutron star systems show a slight excess of power at very high
frequencies, and the third being that kilohertz QPOs are seen in the
neutron stars, but not in the black holes in these states (Sunyaev \&
Revnivtsev 2000; Klein-Wolt 2004).  It has also been shown that the
lags between hard and soft photons are consistent with being the same
in neutron stars and black holes in their low/hard states (Ford et
al. 1999).  It was noted in the abstract of Ford et al. (1999) that
the time lags might present some problem for Comptonization models,
but what was really meant by that statement is that the time lags
might present some problem for models where the time lags are produced
by light travel times through the Comptonizing medium (e.g. Payne
1980).  It has since been shown that it is true that the time delays
cannot be light travel times, but that fact does not prohibit the
radiation mechanism from being Compton scattering (Maccarone, Coppi \&
Poutanen 2000).

These results imply that essentially the same processes are producing
most of the radiation in black holes and neutron stars, and that the
size scales for the emission regions should be linearly proportional
to the mass of the compact object to within a factor of about 1.5.
The other two differences between neutron stars and black holes -- the
extra high frequency power in neutron stars, and the kilohertz QPOs in
neutron stars -- may well be related to surface effects in the neutron
star systems.  The kilohertz QPOs show clear correlations with the
spin frequency of the neutron star, since the separation between the
QPOs is always approximately the spin frequency or half the spin
frequency (van der Klis 2005 and references within), and theoretical
models for these QPOs typically invoke the magnetic field of the
neutron star or some non-uniform radiation pattern from its surface
(e.g. Lamb et al. 1985; Miller, Lamb \& Psaltis 1998).  In any event,
these kHz QPOs, while important probes of the physics of neutron
stars, typically have amplitudes of no more than a few per cent and
are hence energetically not very important.  Sunyaev \& Revnivtsev
(2000) suggested that the excess of high frequency power in the
neutron stars might be related to variability in the emission from the
boundary layer on the surface of the neutron star, although we note
that it might be related to the fact that black holes more efficiently
extract energy from their accretion disks into their jets than do
neutron stars.

\section{Properties of scale invariant jets}
Throughout this paper, we will use the term ``scale-invariant jet'' to
refer to jets whose relevant size scales are all linear with the mass
of the compact object driving the jet.  Given the similarities of the
timing properties of the neutron star and black hole X-ray binaries in
hard spectral states, except that the neutron stars' variability is
faster by a factor very similar to the mass ratio, it seems likely
that the X-ray variability is produced in some scale-invariant manner,
whether in a scale-invariant jet, or a corona which is scale
invariant.  In the next several subsections, we will show that scale
invariant jets are not capable of reproducing the different ratios of
X-ray to radio power seen in neutron stars and black holes.

\subsection{Synchrotron jets}
The properties of synchrotron emission from scale-invariant jets are
the most straightforward of the different possible cases, as the
relevant equations have been worked out in the past and tested
observationally.  It has been found that the mass dependence of the
radio emission is such that $L_R \propto L_X^{0.7} M_{CO}^{0.8}$,
where $L_R$ is the radio luminosity, $L_X$ is the X-ray luminosity,
and $M_{CO}$ is the mass of the compact object (Merloni, Heinz \& Di
Matteo 2003; Falcke, K\"ording \& Markoff 2004).  The mass difference
between black holes and neutron stars would then account for a factor
of about 5 difference in the radio-to-X-ray flux ratio, well below the
observed factor of 30 difference.  If we relax the assumption of scale
invariance in these jets, and allow that the typical size scale can
change independently of the mass, then we can solve for the size scale
of the jets that would be required to match the observations.  Both
the particle energy density and the magnetic field energy density
should scale as $L/R^2$, where $L$ is the total luminosity and $R$ is
the characteristic size scale, for a system in equipartition (or with
a constant ratio of magnetic field strength to equipartition
strength).  The frequency at which a jet becomes optically thin to
synchrotron radiation is given by equation (14) of Heinz \& Sunyaev
(2003) -- $\nu_\tau \propto (R (L/R^2)^{(p+4)/2})^{2/(p+4)}$, which
yields $L R^{-2(p+3)/(p+4)}$, where $p$ is the spectral index of the
electron distribution.  For $p=2.4$, which is required to reproduce
the typical X-ray binary flux spectral index of $\alpha=0.7$, then the
break comes at a frequency which scales as $L R^{-1.68}$. The
deviations from this relation are very small even for rather large
changes in $p$.  The X-ray flux for a given radio flux then scales as
the break frequency to the power of the spectral index, i.e. $(L
R^{-1.68})^{-0.7}$, or $L^{-0.7} R^{1.12}$.  To reproduce the observed
differences between black holes and neutron stars, then, the typical
size scales in the neutron stars would have to be a factor of about 20
smaller than those in black holes at the same luminosity.  This is
inconsistent with the power spectral differences between black holes
and neutron stars which show, at most, a factor of 10 difference
between black holes and neutron stars' variability timescales.

\subsection{Compton scattered jets}

Having established that the fraction of the synchrotron radiation
emitted in the X-rays will not fall off fast enough with mass to allow
that synchrotron X-ray jets provide for the differences between
neutron stars and black holes, let us consider the the possibility
that X-ray jets may be dominated by Compton processes, rather than
synchrotron X-rays.  This allows another free parameter, the ratio
between synchtrotron and Compton fluxes.  

Considering the case of Compton scattered emission, rather than
synchrotron emission does not change the situation. Firstly, in scale
free jets, the ratio of synchrotron to Compton luminosity, is
independent of the compact object mass.  Ghisellini \& Celotti (2001)
show that for synchrotron self Compton jets, the ratio of Compton to
synchrotron luminosity goes as $L_{synch}^{1/2}R^{-2}B^{-2}$, while
$B\propto$$L_{synch}R^{-3}$.  Combining terms shows that the ratio of
Compton to synchrotron luminosity goes as $(L_{synch}/R)^{1/2}$, which
makes this a scale independent quantity.

Neither can this problem be solved by external seed photons for
Comptonization.  Firstly, the only natural source of such seed photons
would be from the neutron star's surface.  The surface emission seems
to be weaker than half the total luminosity in extreme island state
neutron stars (e.g. Piraino et al. 1999; Jonker et al. 2004), which
may be showing that some of the rotational energy that should be
dissipated at the surface of the neutron star is going into the jet,
or may be showing that these photons are being Compton upscattered in
the corona before reaching the observer (i.e. they are simply
providing an extra source of seed photons to a corona with a much
higher optical depth and covering fraction than the jet).  It is clear
that the photons are not being produced in a region which is not
surrounded by the corona; that is, the photons cannot be produced from
the surface of the neutron star, unless the jet's opening angle is
nearly 180 degrees.  The jet may look like a lot a planar wind (see
e.g. Junor, Biretta \& Livio 1999) near the surface of the neutron
star, so this argument alone is not robust.

Secondly, even if there were some other source of such photons, if the
X-rays from X-ray binaries are Compton emission, then the spectrum
should cut off at a much lower photon energy for neutron stars than
for black holes, because the combination of the higher magnetic field
strength and the extra source of seed photons will lead to a
dramatically higher cooling rate.  This cooling rate will lead to a
lower cutoff energy for the electrons, and hence a lower cutoff energy
for the Compton component (see e.g. the discussion of the effects on
inverse Compton emission on the ``blazar sequence'' shown in Fossati
et al. 1998).  A related point is that Compton scattering produces
upscattered photons with energy of $\sim \gamma^2 E_0$, where $\gamma$
is the energy of the electron doing the upscattering, and $E_0$ is
initial energy of the photon.  While the acceleration of electrons may
be governed by processes that lead to the roughly constant cutoff
energy seen in hard state systems, it would require considerable fine
tuning for the same Compton cutoff to be seen when the seed photons
are thermal ($\sim$ 1 keV) photons from the surface of the neutron star
in one case, and synchrotron photons from the jet itself in the other
case.

Finally, let us check the self-consistency of this picture. Taking the
minimum energy requirements from Ghisellini \& Celotti (2001), we find
that the synchrotron luminosity from a jet which is in equipartition,
and where the Compton scattered is dominated by external seed photons
(since this is required to create a difference between neutron stars
and black holes) will be $L_{synch}=
\left(\frac{L_{compt}}{L_{seed}}\right)^2 \left(\frac{\pi R m_e
c^3}{\sigma_T <\gamma^2>}\right).$

This equation assumes the most favorable case for an external Compton
dominated jet, which is a pair-dominated jet.  A self-inconsistency
will be found for the cases such as that of 4U~0614+091, where the
spectrum shows the power law component which is at least 25 times
brighter than the thermal component at a luminosity of about
$4\times10^{36}$ ergs/sec (Piraino et al. 1999).  In this case, the
synchrotron luminosity would have to be at least $6\times10^{35}$ (and
this is assuming an emission region of 10 km size scale, equal to the
neutron star's radius), while the blackbody luminosity would be
2$\times10^{35}$ ergs/sec, about 3 times lower.  Escaping this
constraint requires violating equipartition by a substantial factor
(the exact factor depending on how much bigger the emission region is
than the neutron star).  This may be possible, but would then require
a large total kinetic energy in electrons, which would make the
radiative efficiency even lower than it is normally assumed; since we
show below that the low radiative efficiency of jets is already a
problem for a jet X-ray scenario, this seems an unlikely possibility.

\section{Lack of abrupt luminosity changes at state transitions}
Synchrotron-emitting jets are radiatively inefficient by nature.
Fender (2005) collected a variety of pieces of observational and
theoretical evidence regarding the radiative efficiency of
relativistic jets seen in X-ray binaries, all of which argue for a
radiative efficiency of no greater than 15\% (Blandford \& K\"onigl
1979; Fender \& Pooley 2000; Markoff, Falcke \& Fender 2001).

The outbursts of soft X-ray transients frequently show transitions
between the different spectral states.  It is generally believed that
jet formation in the thin disks of the high/soft state is strongly
suppressed, since jet production requires large scale height magnetic
fields (Livio, Ogilvie \& Pringle 1999). This claim is supported by
the non-detections of radio jets from high/soft state X-ray binaries
(e.g. Tananbaum et al. 1972; Fender et al. 1999).  One exception has
been found, but it is likely that the radio emission in this case was
produced from a plasmon ejected before the transition to the soft
state occurred, but which interaction with the interstellar medium
during the soft state (Corbel et 2004).  Let us assume that the
accretion flow goes from being jet dominated in the X-rays in a hard
state to being in a high/soft state where all the power is radiated
from a geometrically thin disk.  The radiative efficiency of the flow
would then change abruptly from about 15\% or less, in the X-ray jet,
to nearly 100\%, in the thin disk.  We note that this change is in
addition to the fact that jets are unlikely to extract 100\% of energy
from the accretion disks that feed them.  Markoff et al. 2001 estimate
that 0.1-10\% of the disk's energy is extracted into the jet. Given
the most likely parameter values chosen in Markoff et al. (2001), a
jet with 10\% efficiency extracting 1\% of the disk's power, a factor
of $\sim$1000 luminosity increase would be expected across the
low/hard to high/soft state transition, assuming that the mass
accretion rate is changing steadily at that time.

This requirement shows serious disagreement with the observations.
The soft X-ray transients show smooth variations in luminosity, even
across the state transitions, rather than factors of 10 jumps in the
luminosity corresponding to changes in radiative efficiencies of the
accretion flow (see e.g. Sobczak et al. 2000; Miller et al. 2001).  At
the state transitions, sharp changes are seen in the relative flux in
the quasi-thermal and power law components, but no sharp changes are
seen in the bolometric luminosity.  That is, the spectrum
re-configures itself without any dramatic change in luminosity,
indicating that the radiative efficiency is making, at most, a small
change.  We note that this same argument applies to the arguments of
Malzac, Merloni \& Fabian (2004) that the jet should be heavily
dominant over the disk+corona system in order to reproduce the
complication correlations and anti-correlations between the optical
and X-ray emission in XTE~J1118+480.  However, those authors results
were also consistent with a less heavily jet dominated accretion flow,
which would be consistent with no abrupt break in the X-ray luminosity
at the state transition.

Some caveats do exist.  A factor of two change in the radiative
efficiency may be expected at the state transition, because the
rotational energy of the inner edge of the accretion disk should be
advected in the high/soft state.  Some of this rotational energy is
probably supplied to the jet in the low/hard state.  Additionally,
almost half the accretion power is dissipated by the thermal accretion
disk in the highest luminosity low/hard states.  The smallest possible
luminosity change across the state transition for a jet dominated
power law would then be a factor of 2, which is still unlikely given
the observational data, but which cannot be ruled out without
well-sampled soft X-ray through $\gamma$-ray monitoring.

Also, the high/soft state could have a jet at very high Lorentz factor
(see e.g. Meier 1999 who argues that jets powered by thin disks should
be faster than those powered by thick disks; see also Fender, Belloni
\& Gallo 2004 for tentative observational evidence that the softer the
state of the accretion disk from which a jet is ejected, the faster
the jet), which could transport away a large amount of power from the
system.  These jets might be essentially unobservable, because of the
very low probability that the observer would lie within the beaming
cone of the jet.  In any cases, even if such jets do exist, it is
likely that they are substantially weaker than jets powered by
geometrically thick accretion flows.  Searching for the jet-ISM
interaction site in the high/soft state source LMC~X-1 could help test
whether it has a strong but invisible jet.

\section{Discussion}

As we have now established that it is quite difficult to explain the
relevant similarities and differences between neutron stars and black
holes in context of jet models, it is important that we show that a
standard Comptonization model does not present similar qualitative
difficulties in explaining the data.  First, we note that neutron
stars are likely to extract less of their accretion power into jets
than are black holes.  Their potential wells are not as deep, leading
to more slowly rotating orbits around the compact object, and they
spin much more slowly, so they have less spin energy to extract.  As a
result, the two most prominent sources for extracting energy into
jets, the rotational energy of the compact object (Blanford \& Znajek
1977), and that of the accretion disk (Blandford \& Payne 1982) are
suppressed in neutron stars relative to rotating black holes.

Meier (2001), for example, calculated the jet power provided from an
advection dominated accretion flow (see e.g. Narayan \& Yi 1995)
through the Blandford-Znajek (1977) mechanism and the Blandford-Payne
(1982) mechanism.  He found that the jet power should be proportional
to $0.55f^2+1.5fj+j^2$ for black holes, where $j$ is the dimensionless
angular momentum parameter for black holes, and $f$ is a dimensionless
parameter less than 1, which is the ratio between the rotational
velocity of particles in the ADAF calculations of Narayan \& Yi (1995)
and the actual value.  Models of high frequency quasi-periodic
oscillations from black holes typically require $j$ in these systems
to be in the range 0.7-0.95 (e.g. Strohmayer 2001; Abramowicz \&
Kluzniak 2001; Rezzolla et al. 2003), while the accreting neutron
stars rotational velocities are generally in the range 300-600 Hz,
which gives $v/c$ of about 0.05-0.1.  The treatment of a neutron star
as a black hole of the same mass and angular momentum is not strictly
valid, but we are interested here only is showing that reasonable
parameter values can reproduce the observed differences between
neutron stars and black holes.  Evaluating over the bound from 0 to 1
for $f$, this gives a possible range of 3-400 for the ratio between
the jet powers of neutron stars and black holes, assuming that spin is
the only effect, and that the accretion flow feeding the jet is in the
form of an ADAF.  It has also been suggested that magnetic
reconnection powered coronae (see e.g. Haardt \& Mararschi 1993) would
have sufficiently strong poloidal magnetic fields that they could
power strong jets (Merloni \& Fabian 2002).  The dependence of the jet
power on the compact object's spin should be rather similar to that
for the ADAF model calculated by Meier (2001).  It should be noted
that other effects of the difference between black holes and neutron
stars, most notably the neutron stars' magnetic fields and boundary
layers, might also be important.

Next, we speculate that one of the key differences between the Fourier
spectra of black holes and neutron stars might be explained by the
differences in their abilities to extract power from the jet.  Jets in
low/hard state black holes are likely to extract a large fraction of
the total accretion power as kinetic power (e.g. Meier 2001; Fender,
Gallo \& Jonker 2003; Malzac et al. 2004). The fraction must increase
with decreasing accretion rate, so that the faintest sources are
totally jet dominated, but the sources at luminosities just below the
state transition are not losing more than half their power to the jet.
This extraction is most likely to be important from the innermost part
of the accretion flow, where the poloidal magnetic fields are
strongest and the rotational velocities are highest.  This thus means
that at some point, the radiative efficiency in the X-rays should drop
due to the jet's extraction of power, and that this effect should be
more important in black holes than in neutron stars.  Thus the
smallest spatial scales where the fastest variability is seen should
produce a smaller fraction of the power in the black hole systems
relative to the neutron star systems, while there should be no effect
on the relative variability amplitudes in the black holes and neutron
stars at lower frequencies, where the power spectra are observable to
differ only through the mass dependence.

All this is not to say that jets never produce detectable X-rays.  It
is known, for example, that blazars produce synchrotron photons out to
very high energies in some cases (e.g. Pian et al. 1998).  Blazars
also often show the jet luminosity to be dominated by X-ray from the
Compton component of the jet.  In these cases, there is likely to be a
significant contribution of seed photons for Comptonization from the
accretion disk, which is obviously not possible in a system where the
total luminosity is dominated by the jet (Fossati et al. 1998).  In
some AGN, X-ray jets are resolved (Wilson \& Yang 2002; Sambruna et
al. 2004), but in these cases, the size scales on which the resolve
jets are seen are equivalent to the interactions sometimes seen
between jets and the ISM in X-ray binaries; the AGN do not provide a
useful direct test for whether X-ray binaries' X-ray emission comes
from jets.  What seems plausible is that a small fraction of the X-ray
emission, especially in the soft X-rays, does come from the
synchrotron jet in X-ray binaries, but that the jet spectrum is softer
than assumed by past work suggesting that the bulk of the X-ray power
is synchrotron emission.

Standard Comptonization models, rather than synchrotron jet models are
most likely responsible for the bulk of X-ray emission from X-ray
binaries.  Compton corona models are more capable of matching the
different jet properties of black holes and neutron stars. They also
predict continuous variations in luminosity across the state
transitions, as observed, rather than sharp jumps in the luminosity.
Models where the X-ray emission is Compton scattering from a jet are
restricted to a rather small range in parameter space.

\section{Acknowledgments}
I wish to thank Rob Fender, Julien Malzac, Sera Markoff, Andrea
Merloni, Dave Meier, and Simone Migliari for useful discussions; Elena
Gallo both for useful discussions and for reviewing the manuscript;
and the referee for a quick and constructive report.

\label{lastpage} 

\end{document}